\documentstyle[preprint,eqsecnum,aps]{revtex}

 {\par\noindent{\underline{Proof} \quad}}{\hfill$\Box$\bigskip}
 {\par\noindent{\underline{Proof} of the theorem\quad}}{\hfill$\Box$\bigskip}
 {\par\smallskip\noindent{\underline{{\it Remark}} \quad}}{\par\smallskip}
 {\par\smallskip\noindent{\underline{{\it Fact}} \quad}}{\par\smallskip}
 {\par\smallskip\noindent{\underline{{\it Example}} \quad}}{\par\smallskip}
 {\par\smallskip\noindent{{\it Assumotion} \quad}}{\par\smallskip}
 {\par\smallskip\noindent{{\it Condition} \quad}}{\par\smallskip}

\begin{document}
\title{ On the role of coherent attacks in a type of strategic problem related 
to quantum key distribution
}
\author{Wang Xiang-bin\thanks{email: wang$@$qci.jst.go.jp} 
\\
        Imai Quantum Computation and Information project, ERATO, Japan Sci. and Tech. Corp.\\
Daini Hongo White Bldg. 201, 5-28-3, Hongo, Bunkyo, Tokyo 113-0033, Japan}

\maketitle 
\begin{abstract}
{We consider a strategic problem of the Evesdropping to quantum key distribution.
Evesdropper hopes to obtain the maxium information
given the disturbance to the qubits is often 
For this strategy, the optimized individual 
attack have been extensively constructed under various conditions.
However, it seems a difficult task in the case of coherent attack, i.e., Eve may treat a number
of intercepted qubits collectively, including the collective unitary transformations and the
measurements. It was conjectured by  Cirac and Gisin that no coherent attack can be more powerful 
for this strategy for BB84 protocol.
In this paper we give a general conclusion on the role of coherent attacks for
the strategy of maxmizing the information given the disturbance. 
Suppose in  a quantum key distribution(QKD) protocol, all the transmitted bits 
from Alice are independent and only the individual disturbances to each qubits are
examined by Alice and Bob.  For this type of protocols( so far almost 
all QKD protocols belong to this type),
in principle no coherent attack is more powerful than the product of optimized individual
attack to each individual qubits.
All coherent attacks to the above QKD protocols can be disregarded for the strategy
above.   
}\end{abstract}

Since Bennett and Brassard\cite{bennett1} suggested their quantum key
 distribution protocol(BB84 protocol) in
1984, the subject has been extensively studied both theoretically and experimentally. The protocol allows 
two remote parties Alice and Bob to create and share a secret key using a quantum channel and public
authenticated communications. The quantum key created
in this way is in principle secure because eavesdroppers have no way to tap the quantum
channel without disturb it. 
In the protocol, $k$ independent qubits( such as photons) $|Q_A\rangle$ are first prepared by Alice, each one
is randomly chosen from a set of states $V$. In BB84 scheme 
$V=\{|0\rangle,|1\rangle,|\bar 0\rangle, |\bar 1\rangle\}$, $|0\rangle,|1\rangle$ are bases of $Z$ and
 $|\bar 0\rangle, |\bar 1\rangle$ are bases of $X$. According to each individual states, she writes down a string of 
classical bits, $S_{0A}$. She then sends them to Bob. Bob measures each individual
qubits in basis randomly chosen from $Z$ or $X$. For
 those Bob has happened to choose  the correct bases,
the results should be identical to the corresponding bits in $S_{0A}$. Alice and Bob discard the bit values where Bob measures in a wrong basis. After that, Alise has a string $S_A$ and Bob has a string
$S_A'$.  The shared secret key for Alice and Bob can be built up based on this. In the case of
noiseless channel
without Eve., $S_A$ should be identical to $S_A'$ and no third party knows any information about $S_A$.
So far there are many new proposals on QKD scheme. For example, the 6 state protocol\cite{bruss1}, 
the $d-$level qubit protocol\cite{cerf} and so on.

Normally, one can classify Eve's attack into two classes. In an  individual attack,  Eve
operates the qubits from Alice individully  with her ancilla. 
In such an attack, Eve's information to each qubit is independent.
In a coherent attack, Eve may operate a number of qubits from Alice collectively with her ancilla by all possible unitary transformations and measurements.  
In this paper, we give a general proof for the Cirac-Giain conjecture\cite{cirac}. It was shown that,
Eve's total information {\it in average} about the raw bits given a fixed disturbance
does not increase through $2$-qubit coherent attack. No $2-$qubit coherent attack can be more powerful
than the optimized individual attack which maxmizes Eve's total  
 information about the raw key given the disturbance. It is conjectured\cite{cirac} that the conclusion can be also
correct for a coherent attack on arbitrary number of qubits from Alice. But no strict proof has been given there.
So far it is not clear on how Eve's total information about the raw key is connected to the security
of the final key shared by Alice and Bob in the security proofs for the final key\cite{mayers1,mayers2,lo1,lo2,shor}.  However, they could have a relationship by many people's
intuition, since there is indeed a relationship between Eve's information of
raw key and the security of the final key in classical private information .  Cirac-Gisin conjecture could be useful in the future when we are clear of the role
of Eve's information about the raw key.

In our proof, we require that Bob measure every  received qubit independently. This requirement 
is used
 by $all$ QKD protocols proposed so far. However, we don't
add any constraint to Eve in the coherent attack. The quantity of $disturbance$ is measured by the error rate on Bob's measurement result. Here we assume all errors are caused by the channel noise which includes the action of Eve.   We will consider the BB84 protocol in our proof, but the conclusion is obviously
correct for all protocols raised so far.   
 
Most generally, we assume Eve first intercepts state $|Q_A\rangle$ from Alice which includes $n$
qubits, she takes
 a unitary transformation  $\hat U_{AE}$ on both $|Q_A\rangle$ and
her own ancillas state $|E\rangle$. After this transformation, she
sends those quibits originally from Alice to Bob and keeps the ancillas.
Finally she meassures her ancilla to obtain the information about $S_A$ in the future. Here the final measurement $M_E$ can be any type of POVM and not limited
to the projective measurement. Bassically, there are two types
of attack, the individul attack and the coherent attack\cite{ekert2,ekert3,cirac,hwang}. In an induvidual attack, Eve's attack $\hat O$ is the simple product of $ \hat O_1\cdot\hat O_2\cdots \hat O_n $. More prcisely, in an individual measurement, Eve's operation on the qubits can be described by
\begin{eqnarray}
\hat O |Q_A\rangle E\rangle = M_{E1}\cdot U_{AE1}|Q_{A1}\rangle|E_1\rangle \otimes M_{E2}\cdot U_{AE2}|Q_{A2}\rangle|E_2\rangle \otimes \cdots M_{En}\cdot U_{AEn}|Q_{An}\rangle|E_n\rangle 
\end{eqnarray}
Where state $|E_i\rangle$ is the $i'$th ancilla which is attached to the $i'$th
qubit from Alice $|Q_{Ai}\rangle$. $U_{AEi}$ is is the unitary transformation on state $|Q_{Ai}\rangle|E_i\rangle$, $M_{Ei}$ is certain measurement on $i$th
ancilla.
Therefore  an individual attack can be defined as
\begin{eqnarray}
 \hat O_1\otimes\hat O_2\otimes\cdots \hat O_n = M_{E1}U_{AE_1}\otimes
M_{E2}U_{AE2}\otimes\cdots M_{En}U_{AEn}.
\end{eqnarray} 
That is to say, in an individual attack, both the unitary transformation
$U_{AE}$ and the measurement $M_{E}$ are factorizable.
However, in a coherent attack, there is no restriction to either $\hat U_{AE}$ or the measurement $M_E$.
What we shall show is that, given  disturbance to $|Q_A\rangle$, 
it is enough for Eve to use the individul
attack only in order to obtain the maximum amount of information about $S_A$.  
 
Basically, there are two quantities  $I_{EA}$ and $D$ in evaluating the security of a protocol under 
eavesdropper$'$s attack. Here $I_{EA}$ is
the amount of information about   $S_A$ eavesdropper can obtain after the attack,
$D$ is the disturbance to $|Q_A\rangle$. Most generally, 
$D$  can be defined as the distance between $Q_A$ and $\rho_A'$, where $\rho_A'$ is the state sent
to Bob from Eve. 
Here we assume in the QKD protocols  Bob and Alice only examine the individual disturbance to each qubits.
For example, Bob takes the independent measurements to each individual qubits,  each qubits 
have the 
equal probability to be chosen for the check. 
The disturbance is
measured by Bob$'$s error rate of  his measurement results for those qubits
 which are chosen for the check.
So far all QKD protocols work in such a way to estimate the disturbance.
 Therefore the  $detectable$ disturbance  is
dependent on the average disturbances to each individual qubits, $\{D_i\}$.   

Given the protocol and the attacking scheme, the attacking results are in general
different for different initial state $Q_A$.
Here we evaluate an attacking scheme by  the $average$ effect on all possible $|Q_A\rangle$.  
 That is, if eavesdropper chooses the optimized attack over $n$ qubits intercepted from Alice,
 the security is evaluated by the quantities averaged over all possible states for the $n$
independent qubits, each of which are randomly chosen from certain set as required by the specific
QKD protocol itself, and all possible actions( such as the independent measurements) Bob may take
to the $n$ qubits received. 
The ensemble averaged quantity are denoted by 
$\langle I_{EA}\rangle$ and 
$ \{\langle D_i\rangle\}=\{\langle D_1\rangle,\{\langle D_2\rangle\} \cdots\{\langle D_k\rangle \}$  
for a QKD protocol  under ceratin
attack.  

In any attacking scheme, to the eavesdropper the information obtained should be the larger the better
while the disturbance should be the less the better. Given disturbance 
$\{\langle  D_i\rangle\}$, we define the optimized attack 
$\hat O(\{\langle  D_i\rangle\})$ as the one by which 
the eavesdropper  gains the maximum information among all possible
attacking schemes with the same disturbances.\\
{\bf Theorem: No (coherent) attack can be more powerful than the 
optimized individual attack for Eve$'$s strategy of maximizing the total information given the
disturbance.} 
Here we put the word $coherent$ inside brackets because the theorem holds for all
attacks, however, only coherent attacks need a proof.

Specifically, if eavesdropper is  attacking  $m$ qubits which are being
transmitted from  
 Alice to Bob,
she can have two different attacks. One is the coherent attack $\hat O_{cm}$. After
 this attack, the disturbance to the $i$th qubit is $\langle D_i\rangle$. The other is
the product of individual attack 
$\hat O_m=\hat O_0(1)\otimes\hat O_0(2)\otimes\cdots \otimes\hat O_0(m)$. 
After this attack, the disturbance to $i$th qubit is also $\langle D_i\rangle$.
That is to say, to each individual qubit, the individual attack $\hat O_m$ causes the same disturbance
as the coherent attack does. $\hat O_0(i)$ is the optimized individual attack to qubit $i$ given the
disturbance $\langle D_i\rangle$. Explicitly, we give the following definition on optimized
individual attack $\hat O_0(i)$ with fixed disturbance:
\\ {\bf Definition} for the notation $\hat O_0(i)$: 
When  the disturbance $\langle D_i\rangle$ is given,
eavesdropper may obtain the maximum information through $\hat O_0(i)$, among all individual
attacking schemes. There is not any individual attacking scheme by which Eave can obtain more
information about the $i'$s bit than that by scheme $\hat O_0(i)$.  

To show the theorem, we need only show the following Lemma:\\
{\bf Lemma:} No ( coherent) 
attack $\hat O_{cm}$ can help the eavesdropper to obtain more information
 than the individual attacking scheme $\hat O_m$ as defined above.\\
To show this, we use the following idea:\\
{\it Step 1}. When $m=1$ it is obviously correct.\\
{\it Step 2}. Assume the it  is true  when $m=n-1$.\\
{\it Step 3}. We can then prove  it must be also true in the case of $m=n$.

Now we show {\it Step 3} based on the assumption in {\it Step 2} and {\it Step 1}.
We shall do it in the following way:\\
Suppose the phrase in step 3 is {\it not} true. Then there must be a a coherent attack $\hat O_{cn}$
which can outperform the individual attack $\hat O_n$. Here $\hat O_{cn}$ can include 
any collective treatments such as the coherent unitary transformations and the coherent measurments
 to the $n$ qubits Eve. has intercepted from Alice( and Eve$'$s ancilla). 
Then we can construct a game $G$ which is an individual attack to one qubit from Alice. 
After the game $G$ we find $\hat O_0(i)$ is {\it not} the {\it optimized} 
individual attack to a single qubit because it does not work  as effectively as game $G$. 
This conflict shows that the Lemma must be true. 

The game $G$ is played in this way: When Alice and Bob is carrying out the QKD programme, eavesdropper asks
her friends Clare and David do the same QKD protocol. Eavesdropper intercepts 1 qubit from Alice,
and $n-1$ qubits from Clare. Without loss of generality, eavesdropper may put the qubit from Alice
at the $n$th order in this group of  qubits. We denote these $n$ qubits as $|Q_{CA}\rangle$. 
She then carries out the $\hat O_{cn}$ to these $n$ qubits.
Since the first n-1 qubits can be regarded as  ancillas attached to the only qubit from Alice,  $\hat O_{cn}$ here is an individual attack to the qubit from Alice, 
 although it were a coherent
attack if all $n$ qubits had been from Alice.
 After certain operation as required in $\hat O_{cn}$, 
  $|Q_{CA}\rangle$ is changed
into $\rho_{CA}'$ and then she sends the $n$th bit to Bob and the first $n-1$ bits to David. Let David
then play the role as Bob to those $n-1$ qubits. When she completes everything required in $\hat O_{cn}$,
she asks Clare announce the exact information about the first $n-1$ bits. We can show that, with
this announcement, eavesdropper$'$s information to the $n$th bit, i.e., the only Alice$'$s bit,
is larger than that from the individual attack $\hat O_0(n)$. Note that, to the qubit from Alice,
the game $G$ here is an {\bf individual attack}. All $n-1$ qubits from Clare can be regarded as
part of Eve$'$s $ancilla$.   This shows that we can use game $G$, a new individual attack to outperform
the optimized individual attack $\hat O_0(n)$ given the disturbance $\langle D_n\rangle$. This is 
obviously impossible, because we have already assumed that $\hat O_0(n)$ is the optimized individual
attack given $\langle D_i \rangle$.

Now we give the mathematical details of the game $G$ above.
Using  the Shanon entropy\cite{robert,nielson}   we have the following quantity for the (average) 
degree of 
uncertainty corresponding to coherent attack $\hat O_{cn}$:
\begin{eqnarray}
H(\hat O_{cn})= -\overline{\sum_{X, x_n}p(X,x_n)\log p(X, x_n)\label{key}}.
\end{eqnarray} 
$X=x_1,x_2\cdots,x_{n-1}$ and
 $p(X, x_n)$ is eavesdropper$'$s probability distribution for the $n$ bits.
With the definition of $X$, the mathematical symbol $(X,x_n)$ is nothing but $(x_1,x_2,\cdots,x_{n_1},x_n)$. We write it in the form of $(X,x_n)$ because we will play some tricks to $x_n$ latter.
Note that given different input states $|Q_A\rangle$ and different measurement basis taken by Bob,
the same attacking scheme may  lead to different output {\bf Y}. In general Eve$'$s probability 
distribution is dependent on the outcome {\bf Y}. Here the bar over  
$\sum_{X, x_n}p(X,x_n)\log p(X, x_n)\label{key}$ reprents the averaged result over all 
different {\bf Y}. Thus the entropy $H(O_{cn})$ here is the {\bf average} entropy over all possible {\bf Y}.
This bar average, the average over different outcome for {\it one} configuration, is 
{\it different} from the ensemble average, which is the average over {\it different} configurations.
The formula above shows how uncertain Eve is to the $n$ bits after she
 completes her coherent attack $\hat O_{cn}$ 
on the $n$ qubits
( first $n-1$ originally from Clare, the last one from Alice), but before Clare announces the exact
information for his $n-1$ bits.

We also have the (average) entropy  corresponding to the individual attack $\hat O_n$
\begin{eqnarray}
H(\hat O_{n})=\sum_{i=0}^n H(\hat O_0(i))\label{keyi},
\end{eqnarray} 
and
\begin{eqnarray}
H(\hat O_0(i))=-\overline{\sum_x p_i(x)\log p_i(x)},
\end{eqnarray}
$p_i(x)$ is eavesdropper$'$s probability distribution for the value of the $i$th the 
bit through the individual attack $\hat O_0(i)$.
Again, $H(\hat O_0)$ is the average entropy for all possible outcome of one configuration.
The  disturbances casued by the two attacks are
same.
Therefore if the coherent attack $\hat O_{cn}$ here is  more powerful than the 
individual attack $\hat O_n=\hat O_0(1)\otimes \hat O_0(2)
 \cdots \otimes O_0(n) $ 
we must have the following inequality for the ensemble averaged entropy
\begin{eqnarray}
\langle H(\hat O_{cn})\rangle < \langle\sum_{i=0}^n  H(\hat O_0(i))\label{key2}\rangle .
\end{eqnarray}
Now eavesdropper asks Clare announce  bits information to his $n-1$ bits. With the 
exact information
about the first $n-1$ bits, eavesdropper$'$s new average entropy through  game $G$ is
\begin{eqnarray}
\langle H'(\hat O_{cn})\rangle =\langle H(\hat O_{cn})\rangle 
+  \langle\overline{\sum_X p(X)\log p(X)}\rangle. \label{d1}
\end{eqnarray}
Here $p(X)$ is eavesdropper probability ditribution to the first $n-1$ qubits just before 
Clare$'$s announcement, $p(X)=\sum_{x_n}p(X,x_n)$.

Since we have assumed the the Lemma to be true in the case $m=n-1$, i.e., eavesdropper$'$s
information to the first $n-1$ bits through any coherent attack( and also any other attack)
 should never be larger than the information obtained through
the individual attack $\hat O_{n-1}$ before Clare announces the exact results.
Therefore we have
\begin{eqnarray}
-\langle\overline{\sum_X p(X)\log p(X)}\rangle \geq \langle \sum_{i=1}^{n-1}  H(\hat O_0(i))\rangle\label{key3}.
\end{eqnarray}
 Combining this  with the eq(\ref{key2}) and eq(\ref{d1}) 
 we have the following inequality
\begin{eqnarray}
\langle H'(\hat O_{cn})\rangle < \langle H(\hat O_0(n))\rangle\label{con}.
\end{eqnarray}
Now eavesdropper has the exact information to the first $n-1$ bits, $H'(\hat O_{cn})$ 
can also be interpreted as Eve$'$s entropy
of the $n$th bit through game $G$.
The inequality(\ref{con}) shows that eavesdropper$'$s information
on the single bit initially from Alice through game $G$ is larger than her information 
on the same bit through $\hat O_0(n)$. And also we have assumed  the disturbance to that bit
caused by game $G$ is equal to thal caused by $\hat O_0(n)$.
 This is to say, game $G$, which is an $individual$ attack to Alice$'$s qubit,
 can help eavesdropper  to obtain more information to the 
the bit than the optimized
individual attack to the bit with same disturbance. This conflicts with our definition about
$optimized$ individual attack given the disturbance. Thus the 
inequality(\ref{key2})
must be wrong.  Therefore we obtain our theorem.

Thus we draw the following conclusion:
\\ {\it Suppose in in a QKD protocol, all the transmitted bits are independent 
and the measurements are carried out to each individual
qubits independently. To this type of protocols, no coherent attack can be more powerful
than the product of optimized individual attack $\hat O_n$ 
for Eve$'$s strategy of maxmizing the total information given
the disturbance. 
 }\\
Remarks: Our conclusion is only for the raw key stage. With our result, the conclusions in 
ref\cite{bruss1}, ref\cite{cerf} and ref\cite{bruss2}
on Eve$'$s maximum information through individual attack to the $6$ state protocol, $d$-level state protocol and $3$-level state
protocol
are also correct in the coherent attack case. It should be interesting to investigate
the role of coherent attacks with the error correction and privicy amplification being taken into
consideration. We believe our result here can be useful in the future study of this case. 
For example, it is generally believed that a QKD can be unconditionally secure after the error
privicy amplification, given that Eve$'$s information smaller than Bob$'$s information in the raw 
key stage. Our result can greatly simplify the estimation of Eve$'$s maximum information, therefore
the explicit formula as the creteria of security can be obtained easier. In the cases that the raw key 
is directly used for the secrect communication, coherent attacks can be disregarded.

In the end of this paper, we have to clarify something.
Although we have proven the Cirac-Gisin conjeture in a rather general sense, we are not clear on the role
of our conclusion in the most important topic of the optimized Eve$'$s attack towards the final key in the subject
of quantum key distribution. 
On the other hand, here the strategic problem is for the total information, however, the total information at that step is not everything in the whole
game of quantum key distribution\cite{gisinp}. In the strategic problem above,
there is no test for Alice and Bob. This shows we have assumed that the error rate in the test is equivalent to the
disturbance caused by Eve. This is in general not true because there can be a statistical deviation and
only the subset that passes the test will interest the Evesdropper. However, this issue can be resolved
in the case that the number of qubits in the QKD job between Alice and Bob is much larger than the number
of qubits intercepted for the conherent attack, i.e. $k>>n$. This has been illustrated in ref\cite{acin}.  

{\bf Acknowledgement:} I thank Prof Imai Hiroshi for support. I thank N. Gisin, 
 N. Lukenhaus, D. Mayers, Jian-Wei Pan(U. Vienna), Hwang W.Y. and Matsumoto K. 
 for useful discussions. 

\end{document}